# M.K. DAS GUPTA, THE FIRST INDIAN RADIO ASTRONOMER, AND HIS CONNECTION WITH THE 2020 PHYSICS NOBEL PRIZE

ARNAB RAI CHOUDHURI* and RITABAN CHATTERJEE**

*Department of Physics, Indian Institute of Science, Bangalore – 560012. E-mail: arnab@iisc.ac.in.

**Department of Physics, Presidency University, 86/1 College Street, Kolkata – 700073. E-mail: ritaban.physics@presiuniv.ac.in

*Half of the 2020 Nobel Prize is awarded for discovering a super-massive black hole at the centre of our Galaxy. One of the first indications of the existence of a black hole at the centre of a galaxy was found by Jennison and Das Gupta in 1953 while carrying on a radio observation of the source Cygnus A. Mrinal Das Gupta, who was doing his PhD at Manchester University at the time of this discovery, spent the major part of his professional career at Calcutta University. We give an outline of Das Gupta's life and explain the scientific significance of his discovery.*

## Introduction

The 2020 Physics Nobel Prize has been awarded for research on black holes. Soon after Einstein[1] formulated general relativity in 1915, Schwarzschild[2] obtained a solution of the basic equations of the theory describing the structure of space around a point mass. A tantalizing consequence of this solution has intrigued physicists since that time: there can be objects in the astronomical universe within which gravity is so strong that nothing – even light – can escape from them. For about half a century after the formulation of general relativity, the widely held view within the community of physicists was that this is merely a strange theoretical possibility. Such objects, christened 'black holes' by John Wheeler, were suspected not to exist. The 2020 Nobel Prize is an official recognition that black holes are no longer a scientific fantasy, but very much an integral part of the physicist's current view of our Universe.

Half of the Nobel Prize has been given to Roger Penrose, who proved the powerful singularity theorems in the 1960s, along with Stephen Hawking (who was no longer alive to be considered for the 2020 Nobel Prize), which show the inevitability of black holes in the astronomical universe. It has been widely reported by the press in India that there was an Indian prelude to this breakthrough. The Kolkata physicist Amal Kumar Raychaudhuri[3], who taught for many years at what was then Presidency College, arrived at an important equation from the basic principles of general relativity in 1955. The Raychaudhuri equation was the starting point for Penrose and Hawking to prove the singularity theorems. The companion paper by Majumdar in this issue of *Science and Culture* discusses the connection of Raychaudhuri's work with the 2020 Nobel Prize.

The other half of the Nobel Prize has been given to Reinhard Genzel and Andrea Ghez, who demonstrated the existence of a monster black hole having a mass of about $4 \times 10^6$ times the solar mass at the centre of our Galaxy, which is at a distance of about 25,000 light years from us. It is much less known that the work of another Kolkata physicist done in 1953 has connections with this other half of the Nobel Prize. This Kolkata physicist is Mrinal Kumar Das Gupta, who was in the faculty of the Institute of Radio Physics and Electronics, Calcutta University, for many years.

The famous work of Raychaudhuri was done while he was at Kolkata. Das Gupta, however, did his pioneering work when he was doing his PhD on radio astronomy at Manchester University. At that time, radio astronomy was a new science in its infancy. In collaboration with the fellow PhD student Roger Jennison, he made a startling discovery about the nature of the radio source Cygnus A, which was reported in *Nature* in a paper only slightly longer than a page[4]. Nobody realized at that time that this strange discovery could have anything to do with black holes. It is no wonder that the phrase 'black hole' does not occur anywhere in the 1953 paper by Jennison and Das Gupta. It took astronomers nearly two decades to realize that this work was the first radio astronomical discovery indicating the presence of a black hole at the centre of a galaxy.

We mention that Professor Das Gupta was a very active member of Indian Science News Association (ISNA) and served in the Editorial Board of *Science and Culture* for many years. It is very appropriate that we pay our homage to this first internationally renowned Indian radio astronomer in the pages of *Science and Culture*.

*The Early Career of Das Gupta*

We now give an account of the early career of Das Gupta before we come to a discussion of his famous work. Our account of Das Gupta's life is based on the several obituaries which were published at the time of his death[5-7].

By a curious coincidence, the two Kolkata physicists having connections with the 2020 Nobel Prize – Amal Raychaudhuri and Mrinal Das Gupta – were born in the same town in the same year. They were both born in 1923 in the town of Barishal in undivided Bengal (now in Bangladesh). The fathers of both of them were school teachers. As it happens, both of them died also in Kolkata in 2005.

MKDG (as Mrinal Kumar Das Gupta was known to his students and colleagues) obtained his BSc and MSc in physics from Dacca University. A brilliant student, he stood first class first in both these examinations. S.N. Bose of the Bose-Einstein statistics fame was one of his teachers at Dacca University. MKDG as a child dreamed of becoming a scientist and planned to carry on research at Dacca University after the completion of MSc. But the uncertain conditions of the era charted out a different path for him. The Indian Independence and the Partition came almost immediately after the completion of his MSc. The Das Gupta family decided to relocate to Kolkata. It is not difficult to imagine that MKDG, as the son of a middle class family, felt a pressure to find a job and have an earning.

While MKDG was searching for a job, one day he came across a newspaper advertisement. Professor Sisir Kumar Mitra of Calcutta University was looking for an assistant. Mitra was an internationally acknowledged authority on the upper atmosphere and the ionosphere. When MKDG was a high school student in Dacca, he had once come to Dacca to give a public lecture. The young MKDG was fascinated by the lecture. With his application in hand, MKDG went to meet Mitra in his office in the Rajabazar Science College campus of Calcutta University. He was lucky to be selected for the job.

MKDG[8] had written a charming reminiscence of Mitra, describing his personal association and interactions with Mitra. By the efficient execution of any job that Mitra would entrust on him, MKDG quickly became a trusted confidant of Mitra. At that time, research in the new field of radio astronomy had begun only in a few universities around the world. The far-sighted Mitra realized that this new field was soon going to be very important. He wanted that somebody in his research group should get hands-on training in radio astronomy and should initiate research in this field. Who could be better suited for this job than the young Mrinal? Due to Mitra's endeavours, a scholarship could be arranged for Mrinal to do PhD at Manchester University.

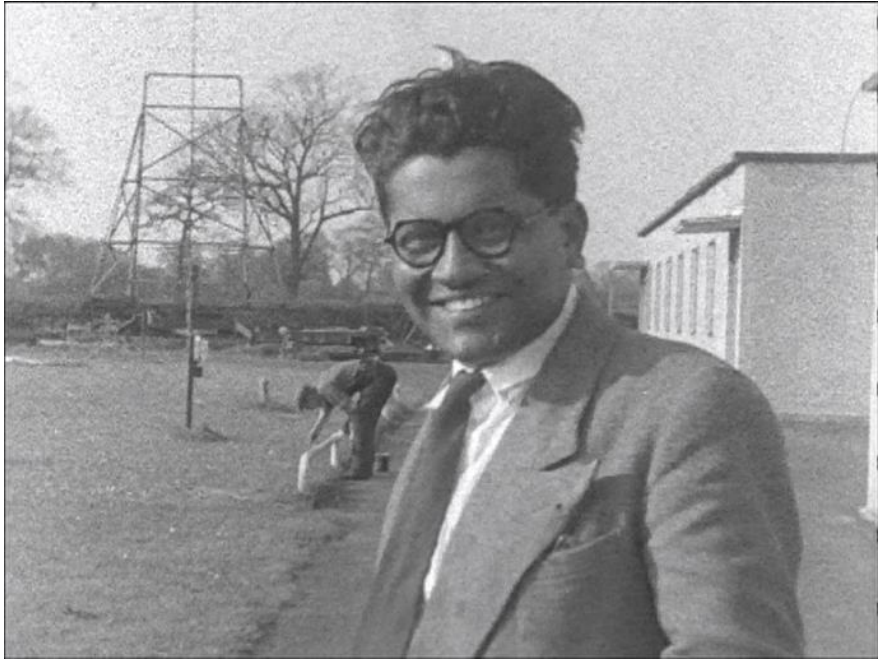

**Figure 1.** Young Mrinal Das Gupta at the Jodrell Bank Observatory. Credit: Jodrell Bank archives.

The Jodrell Bank Laboratory at Manchester University was one of the most important centres of radio astronomy at that time (and continues to remain so till today). Mrinal sailed to Manchester in 1950 and got an opportunity of working for his PhD under the supervision of the legendary radio astronomer Hanbury Brown. He became a close friend of Roger Jennison, another PhD student of Hanbury Brown. These two friends together made the totally unexpected discovery for which nobody in the astronomical world of that time was prepared.

*An Unexpected Discovery*

It was the very early days of radio astronomy. Several celestial radio sources had been discovered but their distance, morphology or even their precise location on the sky was not known. Jennison and Das Gupta used a newly discovered technique, termed interferometry, to observe such a radio source named Cygnus A. The ability of a telescope to precisely determine the location of a source on the sky or its internal structure depends on the telescope's angular resolution. The angular resolution $\theta$ of a telescope using electromagnetic radiation of wavelength $\lambda$ and having a diameter $D$ of the aperture is given by

$$\theta \approx \frac{\lambda}{D}. \tag{1}$$

Clearly, we have to make the diameter larger and the wavelength shorter to resolve smaller angles. Since the wavelength of radio waves is much longer than that of visible light, to attain a resolution comparable to that of a one-meter optical telescope, a radio telescope must have a diameter as large as 10 km! Instead of building single radio receivers of gigantic size, astronomers use a clever technique named radio interferometry – developed by Sir Martin Ryle who went on to win the Nobel Prize for it in 1974 – to achieve the same. In its simplest form, two small radio telescopes located, say 10 km apart, are used for observing the same source at the same time. The signals thus received by the telescopes are then combined in a specialized manner to obtain an image of the source that is equivalent, in many aspects, to what would be produced by a telescope of diameter 10 km.

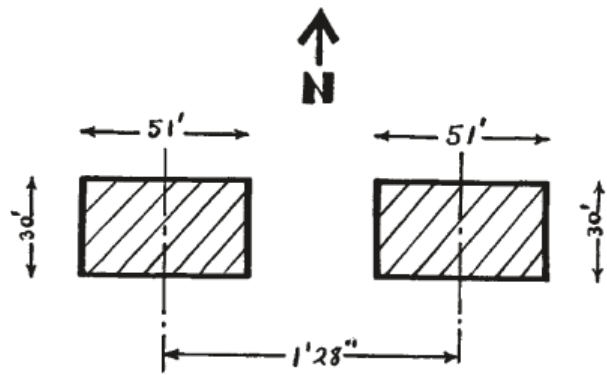

**Figure 2.** Image of Cygnus A obtained by Jennison and Das Gupta[4]. It shows that Cygnus A is not a single source of radio emission but two distinct sources with a gap between them. Note that there were typographical errors in the sizes of the source given in the figure: 51" is written as 51' and 30" is written as 30'.

Using this technique, Jennsion and Das Gupta carried on a study of the radio source Cygnus A. This work was one of the first applications of the concept of intensity interferometry, of which the theory was being developed by Hanbury Brown and Twiss at the same time[9]. Cygnus A, which was previously thought to be a single source of radio waves, was found to consist of two distinct sources with a wide gap between them[4]. Figure 2 shows the radio image presented in the Jennison and Das Gupta paper published in *Nature*. This totally unexpected discovery created a stir in the world of astrophysics. Nobody had any idea how to explain this strange morphology of Cygnus A.

Approximately at the same time, Walter Baade and Rudolph Minkowski[10] identified a galaxy at the position of the radio source Cygnus A, by carrying on observations at the optical wavelength. Interestingly, the galaxy was located right at the centre of the two discrete radio sources discovered by Jennison & Das Gupta. From the optical observation, it was also possible to determine the distance of the source to be 600 million light years, which in turn implied that the two distinct radio sources are 150 thousand light years apart and are emitting as much energy as a hundred billion Suns. During the 1950s, several celestial radio sources were identified to be similar to Cygnus A, i.e., consisting of two distinct sources, termed *radio lobes*. How such an enormous amount of energy is being emitted so far from the centre of the galaxy became one of the foremost outstanding questions of astronomy of that era. It took astrophysicists almost two decades to arrive at a satisfactory theoretical explanation and to realize that these double radio sources could have anything to do with black holes.

*Black Holes as Central Engines of Active Galaxies*

The visible light from a normal galaxy mainly comes from the stars present in it. In 1943 Carl Seyfert discovered that some galaxies had unusually bright centres[11]. This was the first indication that galactic emission may contain something other than the light from its stars. In particular, there may be something unusual at the centres of some galaxies. Such galaxies are known as *active galaxies*. The most extreme examples of active galaxies are what are known as *quasar*s. They often look like very bright stars, but their spectra are very different from the spectra of stars. Ordinary stars have many absorption lines in their spectra, while quasar spectra are dominated by emission lines. Furthermore, the width of the emission lines indicated that they were being generated in matter that is moving at speeds as large as 10,000 km/s. Such high speeds are not at all expected in the stellar atmosphere. In the early 1960s the physical nature of quasars was a topic of intense debate among astronomers.

Finally, in 1963, Marten Schmidt was able to explain the nature of quasar spectra[12] and in the process realized that they were located billions of light years away from us. Quasars were the most distant objects in the Universe ever observed by humans. Another startling implication of this was that quasars must have enormous intrinsic luminosity if they appear so bright from such a large distance: a typical quasar is as bright as a hundred billion suns. In addition, astronomers also inferred from some other observations that the energy-emitting central regions inside quasars were not any larger than our solar system. How can such enormous luminosity be emitted from such a small region?

Many astrophysicists around the middle of the twentieth century thought that black holes can never be detected by our telescopes even if they exist, since light cannot come out of a black hole. However, in the 1960s several theoretical astrophysicists independently realized that all the unusual properties of quasars and other active galaxies can be explained by assuming the existence of a black hole having a million to a billion times solar mass at the centre[13-15]. When matter near the central region of the galaxy gets attracted and eventually falls into the black hole, the matter loses its gravitational potential energy and a part of that lost energy may be emitted in the form of intense radiation before the falling matter disappears into the black hole. The presence of such a *super-massive black hole* (SMBH) implies that gas clouds near the galactic centre would revolve around this black hole at speed of order 1,000–10,000 km/s, causing the broadening of the emission lines. This explanation of the high luminosity of the central region of active galaxies and their broad emission line spectra are now well accepted in the astronomical community.

*From the Central Black Hole to Radio Lobes*

At first sight, it may appear that the discovery of Jennison and Das Gupta has nothing to do with a black hole at the centre of a galaxy. After all, they observed two radio lobes outside the galaxy and not anything at the centre of it. We now come to the discussion why this work should be regarded as the second important landmark (after Seyfert's 1943 paper) in the exploration of black holes at galactic centres.

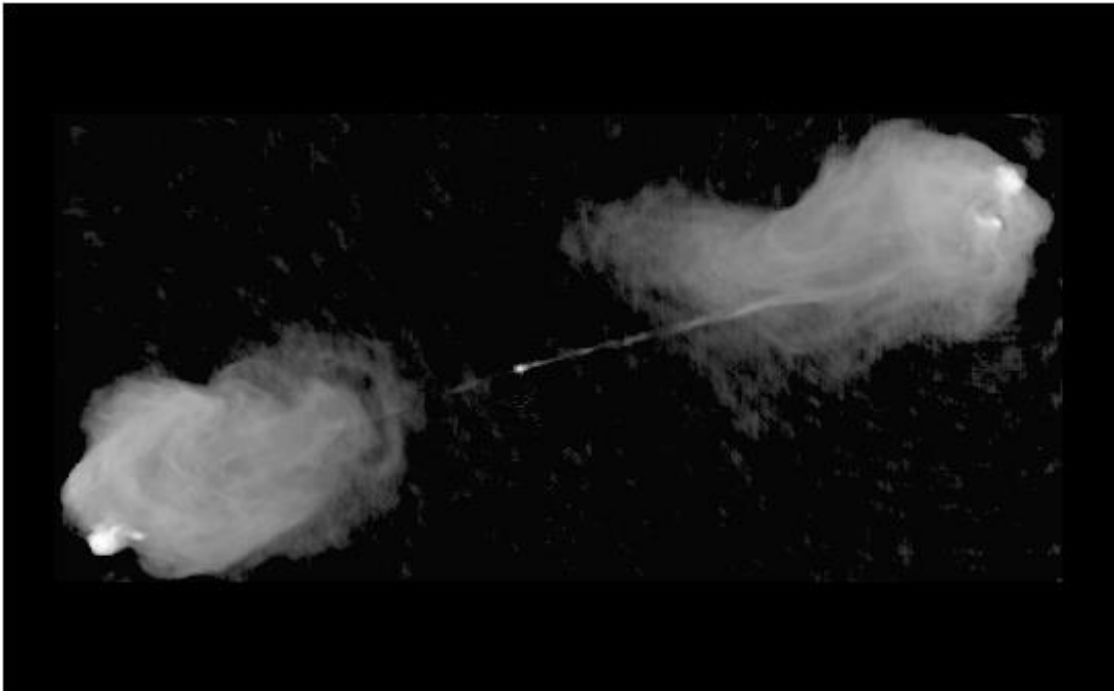

**Figure 3**. Image of Cygnus A obtained with Very Large Array, one of the best radio interferometers at work currently. It clearly shows jet like structures extending from the centre of the galaxy to the lobes. Credit: National Radio Astronomy Observatory.

It is said that a picture says a thousand words. We now show an image of Cygnus A taken with one of the best radio telescopes at work today, using the interferometry technique. Comparing this Figure 3 with Figure 2, we at once appreciate the tremendous advances made in radio astronomical techniques since the time of Jennison and Das Gupta. It can be seen in Figure 3 that very narrow and long structures have extended from the centre of the galaxy on both sides. These are termed *radio jets*. The radio lobes are created where these jets end. It is not difficult to assume that jets are actually outflows of matter that also carries energy from the centre to the radio lobes, from where the energy is radiated in the form of radio waves. In the early 1970s, Blandford and Rees did theoretical calculations to show that such jet-like structures may carry energy to the lobes[16-17].

The big remaining question is: how are the jets produced? The black holes at galactic centres are typically rotating very fast. When matter falls onto a rotating black hole surrounded by magnetic fields, a small amount of the accreting matter may be thrown out at very high speed from the polar regions of the black hole. Detailed calculations have demonstrated this[18-19]. It is now well accepted in the astronomy community that the jets and radio lobes arise from the central supermassive black hole at the radio galactic centre. In hindsight, we realize that the double lobe structure of Cygnus A discovered by Jennison and Das Gupta was the first radio observation indicating the existence of a black hole at a galactic centre.

A careful look at Figure 3 shows a blob at the centre of Cygnus A. In other words, the angular resolution is not high enough to show the black hole inside this blob. Due to tremendous advances in radio interferometry, radio astronomers are now able to combine signals from different radio telescopes existing all over the Earth. In observations done in this manner, $D$ appearing in (1) is essentially the diameter of the Earth, allowing astronomers to resolve very small angles. A project based on this principle – the *Event Horizon Telescope* project – has at last been able to produce recently a striking image (Figure 4) of a black hole at the centre of a galaxy known as M 87[20].

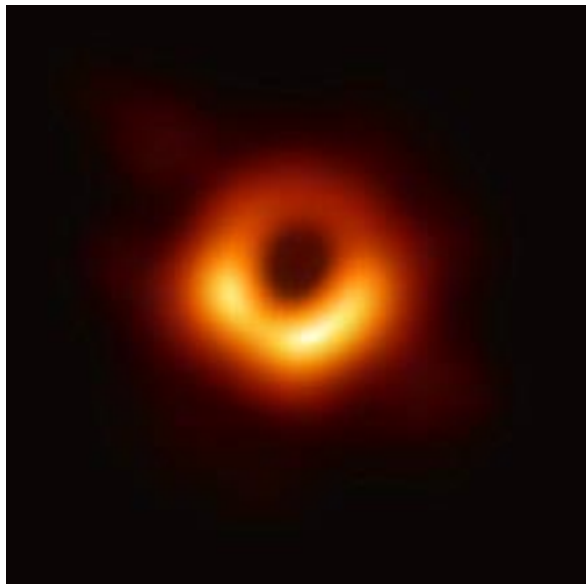

**Figure 4.** The black hole at the centre of the galaxy M 87 imaged by the Event Horizon Telescope. Credit: Official website of Event Horizon Telescope.

*A Black Hole at the Centre of Our Galaxy*

Starting from the early 1970s, astrophysicists have discussed the possibility that, apart from active galaxies, ordinary galaxies also may contain super-massive black holes (SMBH) at their centres[15]. In that case, does our own Galaxy, Milky Way, also contain such an SMBH? If that is the case, then a star in the central region of our Galaxy should move in a Keplerian orbit of elliptical shape around this black hole. The best way of settling the question is to accurately monitor motions of stars in the central region of our Galaxy. The concentration of dust particles, which obstruct visible light, makes the galactic centre inaccessible in visible light. However, the near-infrared emission from the stars there can pass through this veil of dust, making it possible for astronomers of observe stars at the galactic centre in near-infrared wavelengths. Extremely high angular resolution is needed to study the motions of these stars. Such studies can be carried out only with the largest telescopes.

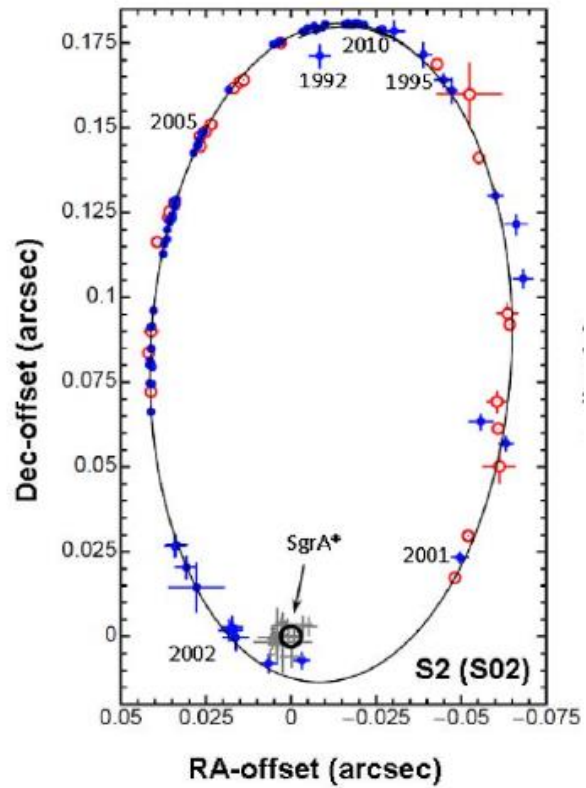

**Figure 5.** The orbit of the star S2 (S02) during the era 1995 – 2010. The small filled circles in blue correspond to the data of Genzel's group, whereas the bigger open circles in red correspond to the data of Ghez's group. Credit: Official website of the Nobel Prize[21].

The groups of Ghez and Genzel studied motions of stars at the galactic centre in the near infra-red by using two of the world's largest telescopes: the Keck Telescope in Hawaii (Ghez) and the Very Large Telescope operated by the European Southern Observatory in Chile (Genzel). The images of astronomical objects get blurred by air turbulence in the Earth's atmosphere. Astronomers have developed two techniques for correcting this: speckle imaging and adaptive optics[21]. Although the groups of Ghez and Genzel started their studies of the motions of stars at the galactic centre from the mid-1990s[22-23], the positions of these stars could be measured really accurately when these techniques were fully developed and implemented a few years later[24-25]. A particular star, called S2 by Genzel's group and S02 by Ghez's group, was found to be especially convenient, since it has a period of about 16 years for going around the galactic centre. The orbit of the star mapped out by the two groups over several years agreed remarkably well[26-27], as shown in Figure 5. The mapping of the orbit of this star led to a good estimate of the mass of the black hole at the centre of our Galaxy, which was found to be about $4 \times 10^6$ times solar mass. Also, the estimates of the size of the central mass from the stars which come very near it indicated that it is unlikely to be anything else except a black hole. Although the orbit of the star S2 subtends a fantastically small angle of only about 0.2 arc seconds in the sky and needs very large telescopes to study it, the closest distance of this star from the black is about 1400 times the Schwarzschild radius. Hence the general relativistic corrections to the orbit are negligible (see, for example, Choudhuri[28], Section 13.3.1).

*Black Holes formed from Dead Stars*

We have primarily restricted our discussion to black holes at galactic centres, since the work of Jennison and Das Gupta was one of the first indications for black holes of this type. For the sake of completeness, we very briefly discuss another type of black holes known to astronomers.

Nuclear reactions take place at the centres of normal stars and the heat produced in this process causes a high pressure at the stellar centre, which balances the radially inward gravitational attraction (see, for example, Choudhuri[28], Chapter 3). When the nuclear fuel is exhausted at the stellar core, the inward gravitational attraction may not be balanced in this manner and the star may start contracting. A quantum mechanical effect – the *degeneracy pressure* – was initially expected to balance the gravitational attraction after the star has contracted sufficiently. Subrahmanyan Chandrasekhar showed in the 1930s that balancing the gravitational attraction by the degeneracy pressure is possible only if the mass of the star is less than a mass limit[29]. Because of our lack of knowledge of matter at this superdense state, the mass limit beyond which the degeneracy pressure cannot balance the inward gravitational attraction is not accurately known. Best estimates of this maximum mass limit suggests it to be in the range 2–3 solar masses. Stars more massive than this may keep on contracting due to their own gravitational attraction once the nuclear burning process stops – leading to the formation of black holes. Chandrasekhar's work was the first indication that black holes may indeed form in the astronomical universe. Most astronomers of that era did not believe in this possibility and Chandrasekhar was awarded a Nobel Prize only in 1983 about half century after this famous work.

In the 1960s, astronomers finally found observational evidence that some dead stars indeed become black holes. If the dead star is in a binary stellar system, then it can attract matter from the outer layers of its companion star. The matter accreting onto the black hole in this manner loses gravitational potential energy, of which a part can be radiated away in the form of intense radiation before the matter falls into the black hole – essentially the same process which occurs around black holes at galactic centres. Theoretical calculations suggest that the radiation produced in this manner will be primarily in X-rays. Since our upper atmosphere is opaque to X-rays, astronomers can detect X-rays from astronomical sources only by flying X-ray detecting instruments in satellites revolving around the Earth above the atmosphere. X-ray observations from satellites done in the 1960s showed X-rays coming from binary stars, vindicating the idea that sufficiently massive dead stars become black holes. Ricardo Giacconi, the pioneer of X-ray astronomy who led these observational efforts[30], was awarded the Physics Nobel Prize in 2002.

The first detection of gravitational wave, for which the 2016 Physics Nobel Prize was awarded to Rainer Weiss, Barry Barish and Kip Thorne, involved the coalescence of two black holes of masses 36 and 29 times the solar mass[31]. Presumably, such black holes formed from the collapse of very massive stars.

*Later life and a Controversy*

After returning with a PhD degree from Manchester, MKDG was promoted to the position of assistant professor at Calcutta University from the position of assistant to S.K. Mitra. He remained with Calcutta University till his retirement. He quickly built up his reputation among the students as an outstanding teacher. He was also very much interested in science outreach and was in great demand to give public lectures on popular science. However, the original expectation of Mitra that somebody from his group should establish a strong research group in radio astronomy after receiving training at the cutting edge research frontiers was not fulfilled. MKDG did keep publishing a few papers on solar radio astronomy in standard journals at a rather low rate and supervised a few PhD theses[7]. However, his research during his long tenure at Calcutta University made little impact in the field, as can be judged by the fact that his few papers published from Kolkata hardly received any citations.

About a decade after MKDG's return from Manchester, Govind Swarup gave up the position of assistant professor at Stanford University and joined Tata Institute of Fundamental Research (TIFR) with the dream of building a world class radio astronomy group in India[32-33]. Within a few years, his group placed India in an honourable position in the world map of radio astronomy. It may be mentioned that radio astronomy research of international standard required considerable funds, which were quite vast by the standards of that time. It was perhaps not possible for MKDG to obtain the kinds of funds which Homi Bhabha could find for Swarup at TIFR. However, when MKDG was doing his PhD at Manchester, some of the Indian science leaders were already aware of the growing importance of radio astronomy. MKDG has written[34]: "The author particularly remembers the visits of Bhabha, Saha and Sarabhai to the Jodrell Bank Experimental Station, Manchester, in the early fifties when he had been working there. They were very much impressed and perhaps were mentally prepared to see this new science emerges in India in due course." S. Ananthakrishnan, who was a research scholar of Swarup and helped him in building the famous Ooty Radio Telescope, had his first training in electronics as a student of Calcutta University. He still declares that he was inspired by MKDG to choose radio astronomy for his career. It is somewhat of a puzzle to us why somebody like MKDG, who had done such an important work during his PhD in Manchester and who could inspire students in this manner, did not establish a good research group in radio astronomy in his home turf. In a short account of the history of radio astronomy in India, MKDG[34] graciously and generously acknowledged the contributions of Swarup and other younger scientists, admitting honestly that his own work done in India was not significant.

We cannot resist the temptation of pointing out that MKDG's career has a remarkable parallel with the career of an exact contemporary: S.N. Ghoshal[35] (1923 – 2007). While doing his PhD at Berkeley under the supervision of Emilio Segre, Ghoshal carried out a famous experiment to verify the compound nucleus model[36]. He returned to India after his PhD and spent several years at Presidency College, Kolkata. He was also not very active in research after returning to

India. Like MKDG, Ghoshal was also known as an outstanding teacher. Ghoshal wrote several excellent undergraduate textbooks, which are still widely used by Indian students. A historian of Indian science should ponder over the question why such obviously brilliant persons like Das Gupta and Ghoshal did not carry on any significant research after returning to India following their famous works done during their PhD years abroad.

Unfortunately, the image of MKDG was badly tarnished in later years due to a controversy. On 3 October 1978, the world's second baby conceived by in vitro fertilization ('test tube baby') was born in Kolkata due to the efforts of Subhas Mukhopadhyay – only 67 days after the birth of the world's first IVF baby, for which Robert G. Edwards was awarded the 2010 Nobel Prize in Physiology or Medicine[37-38]. Since there was widespread scepticism within the Kolkata medical community about the authenticity of Mukhopadhyay's claims, a committee was formed by the West Bengal Government to evaluate this work. Surprisingly, it was chaired by MKDG, who had no professional competence in reproductive biology to evaluate such a work. Although the full report of the Committee was never made public, it is widely believed that the report was not favourable to Mukhopadhyay. He subsequently faced various kinds of official harassment (such as denial of permission to attend international conferences to which he was invited and transfer to an ophthalmology department), tragically ending his own life on 19 July 1981. His extraordinary scientific achievement is now fully acknowledged posthumously[39].

It is alleged in certain circles that the Das Gupta Committee and MKDG as its chairman should be held responsible for Mukhopadhyay's suicide. There is no question that it was an unpardonable professional transgression on the part of MKDG to agree to chair a Committee for which he had no technical qualification. However, due to the non-availability of the Committee report and other relevant documents, the role played by MKDG in the Committee remains unclear. The main known recommendation of the Committee was that Mukhopadhyay should immediately publish his research results in peer-reviewed international medical journals, which he had not done. The scientific breakthrough leading to the birth of the first IVF baby was reported promptly in a top medical journal[40]. The Das Gupta Committee met on 18 November 1978 and must have submitted its report in the next few days, since Mukhopadhyay wrote a letter to the West Bengal Government on 1 December 1978 in response to the deliberations of the Committee[37]. Mukhopadhyay's suicide took place more than two-and-half years after the term of the Committee was over. There is also absolutely no evidence that this Committee should be held responsible in any way for the subsequent difficulty or even harassments which Mukhopadhyay faced from his colleagues and superiors. As T.C. Anand Kumar concludes[37], 'ignorance of the medical fraternity, bureaucratic arrogance and vindictiveness' were the primary drivers behind Mukhopadhyay's suicide.

We point out one fact which appears significant to us. From several persons close to MKDG (including Prasanta Basu, who was MKDG's colleague for many years), we have found that MKDG never discussed his role in the Committee to evaluate Mukhopadhyay even to his most intimate friends. Only 16 long years after Mukhopadhyay's suicide, the veracity of his claims was established when T.C. Anand Kumar examined his laboratory records[37] (in 1997). During these 16 years, it was widely believed that Mukhopadhyay's claims were false and there was no reason for MKDG to be silent on this matter. Why did he still never discuss this case even with his closest friends during this era? What might be going through his mind? We leave these questions for the reader to ponder over.

*Concluding Remarks*

Although the image of Cygnus A obtained by Jennison and Das Gupta has been superseded by incomparably superior images made possible due to the advances in radio astronomy and their paper may not be of much contemporary relevance, it is undoubtedly a landmark step in the development of the field. Their paper, along with an earlier paper[41] explaining the method used in this work, were selected as two of the classics of radio astronomy by Sullivan[42]. The work of Jennison and Das Gupta[4] is also highlighted prominently in the history of radio jets by Spencer[43]. In the famous review article on extragalactic radio sources by Begelman, Blandford and Rees[44], the section on observations begins with a discussion of the paper by Jennison and Das Gupta[4]. Unfortunately, not all scientists are so conscious of their historical heritage and this paper is not cited in the present-day literature of the subject as often as it should be. Perhaps one reason why this landmark work is not remembered so well in the scientific community is that neither Jennison nor Das Gupta was too active in research in later life. Had they remained well-known contributors to radio astronomy, then their important paper might have been remembered better.

The radio interferometry technique, of which the map of Cygnus A produced by Jennison and Das Gupta was one of the first spectacular demonstrations, has become more and more sophisticated. Square Kilometre Array (SKA), the world's largest radio telescope under construction, will be based on this technique. The headquarters of SKA are located at Jodrell Bank, where Jennison and Das Gupta carried out their pioneering work. We would like to mention that several Indian radio astronomers are involved with the SKA project. Another recent international recognition for the Indian radio astronomy community is that the Giant Metrewave Radio Telescope (GMRT), built near Pune in the 1990s under the leadership of Swarup, has been accorded the prestigious IEEE Milestone status. We end by reminding the readers that Mrinal Das Gupta was the first Indian radio astronomer to achieve international renown.

*Acknowledgements*

We are grateful to Suprakash Roy for urging us to write this paper and to S. Ananthakrishnan and Suchetana Chatterjee for valuable discussions. The controversy involving Subhas Mukhopadhyay was brought to our attention by Pratap Raychaudhuri. We thank Prasanta Basu and Malay Chatterjee for sharing their insights about this controversy with us.

*References*

1.  A. Einstein, *Annalen der Physik*, **49**, 769 (1916).
2.  K. Schwarzschild, *Sitzungsberichte der Königlich-Preussischen Akademi der Wissenschaften,* 189 (1916).
3.  A.K. Raychaudhuri, *Phys. Rev.* **98**, 1123 (1955).
4.  R.C. Jennison and M.K. Das Gupta, *Nature* **172**, 996 (1953).
5.  A.K. Datta, *Science and Culture* **72**, 84 (2006).

6. S. Ananthakrishnan and P.K. Basu, *Current Science* **90**, 1288 (2006).
7. S. K. Pal and P.K. Basu, *Biographical Memoirs of Fellows of INSA* **30**, 209 (2006).
8. M. K. Das Gupta, *Resonance* **5**, Issue 7, 92 (2000).
9. R. Hanbury Brown and R.Q. Twiss, *Phil. Mag. A* **45**, 663 (1954).
10. W. Baade and R. Minkowski, *Astrophys. J.* **119**, 215 (1954).
11. C. Seyfert, *Astrophys. J.* **97**, 28 (1943).
12. M. Schmidt, *Nature* **197**, 1040 (1963).
13. Ya. B. Zel'Dovich and I. D. Novikov, Sov. *Phys. Dokl.* **9**, 246 (1964).
14. E. E. Salpeter, *Astrophys. J.* **140**, 796 (1964).
15. D. Lynden-Bell, *Nature* **223**, 690 (1969).
16. M. J. Rees, *Nature* **229**, 312 (1971)
17. R.D. Blandford and M. J. Rees, *Mon. Not. R. Astron. Soc*. **169**, 395 (1974).
18. R.D. Blandford and R.L. Znajek, *Mon. Not. R. Astron. Soc.* **179**, 433 (1977).
19. R. D. Blandford and D.G. Payne, *Mon. Not. R. Astron. Soc.* **199**, 883 (1982).
20. The Event Horizon Telescope Collaboration, *Astrophys. J. Lett.* **875**, L1 (2019).
21. The Nobel Committee for Physics, https://www.nobelprize.org/uploads/2020/10/advanced-physicsprize2020.pdf (2020).
22. A. Eckart and R. Genzel, *Mon. Not. R. Astron. Soc.* **284**, 576 (1997).
23. A. M. Ghez, B. L. Klein, M. Morris, and E. E. Becklin, *Astrophys. J*. **509**, 678 (1998).
24. R. Schödel, T. Ott, R. Genzel, et al., *Nature*, **419**, 694 (2002).
25. A. M. Ghez, G. Duchene, K. Matthews, et al., *Astrophys. J.* **586**, L127 (2003).
26. A. M. Ghez, S. Salim, N. N. Weinberg, et al., *Astrophys. J.* **689**, 1044 (2008).
27. S. Gillessen, F. Eisenhauer, T. K. Fritz, et al., *Astrophys. J*. **707**, L114 (2009).
28. A.R. Choudhuri, Astrophysics for Physicists, Cambridge University Press, (2010).
29. S. Chandrasekhar, *Mon. Not. R. Astron. Soc.* **295**, 207 (1935).
30. R. Giacconi, H. Gursky, F.R. Paolini and B.B. Rossi, *Phys. Rev. Lett*. **9**, 439 (1962).
31. B.P. Abbott, R. Abbott, T.D. Abbott, et al., *Phys. Rev. Lett*. **116**, 241103 (2016).
32. G. Srinivasan, *Current Science* **109**, 618 (2015).
33. W. Orchiston and S. Phakatkar, *Journal of Astronomical History & Heritage* **22**, 3 (2019).
34. M.K. Das Gupta, *Indian J. of Radio & Space Phys.* **19**, 484 (1990).
35. A.N. Saxena, *Physics Today*, https://physicstoday.scitation.org/do/10.1063/PT.4.2133/full/ (2007).
36. S.N. Ghoshal, *Phy. Rev*. **80**, 939 (1950).
37. T.C. Anand Kumar, *Current Science* **72**, 526 (1997).
38. A. Bharadwaj, *Reproductive BioMedicine and Society Online* **2**, 54 (2016).
39. B. Benderly, *Science,* https://blogs.sciencemag.org/sciencecareers/2011/01/the-award-of-th.html (2011).
40. P.C. Steptoe and R.G. Edwards, *Lancet* **2**, 366 (1978).
41. R. Hanbury Brown, R.C. Jennison and M.K. Das Gupta, *Nature* **170**, 1061 (1952).
42. W.T. Sullivan III, Classics in Radio Astronomy, D. Reidel Publishing Co. (1982).
43. R, Spencer, *Galaxies* **5**, 68 (2017).
44. M.C. Begelman, R.D. Blandford and M.J. Rees, *Rev. Mod. Phys*. **56**, 255 (1984).